\documentclass[global,twocolumn,hyperref]{svjour-arXiv}

\usepackage{graphics}
\usepackage{here}
\usepackage{textgreek}
\usepackage{color}
\usepackage{cite}

\begin{document}
\title{Observation of Spider Silk by Femtosecond Pulse Laser Second Harmonic Generation Microscopy\\
}
\author{\and{Yue Zhao}\inst{1}\hbox{\href{http://orcid.org/0000-0002-8550-2020}{\includegraphics{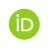}}} 
\and Yanrong Li\inst{1} \and Khuat Thi Thu Hien\inst{1} 
\and Goro Mizutani\inst{1}\thanks{\emph{Goro Mizutani:} mizutani@jaist.ac.jp}\hbox{\href{http://orcid.org/0000-0002-4534-9359}{\includegraphics{orcid.eps}}}
\and Harvey N. Rutt\inst{2}
}                     
%
%

\institute{School of Materials Science, Japan Advanced Institute of Science and Technology, 1-1 Asahidai, Nomi, 923-1292, Japan \and School of Electronic and Computer Science, University of Southampton, SO17 1BJ, UK}

\date{
\\
\\
{\it Surf. Interface Anal.}, {\bf 51,} 56-60 (2019).  DOI:10.1002/sia.6545}

\maketitle
\begin{abstract}
\bf An asymmetric \textbeta -sheet structure of spider silk is said to induce optical second harmonic generation. In this paper, using an in-house non-scanning type femtosecond pulse laser second harmonic generation microscope, we characterized the behavior of the \textbeta -sheet of spider silk under an applied external force. The orientation of the \textbeta -sheets was more unidirectional when the silk was extended. One of the origins of the high mechanical strength of the dragline is suggested to be the physical arrangement of its \textbeta -sheets.
\end{abstract}
\section{Introduction}\label{Introduction}

Spider silk has always been an intriguing material, with a unique combination of properties. It is important to analyze the molecular configuration of the spider silk. Traditional analytical methods are NMR, micro-Raman spectroscopy and micro X-ray diffraction, etc.  Helical and \textbeta -sheet structure of spider silk were reported \cite{g3-30-13,g3-31,g3-20,g12,g3-30,g12-41,g12-39}, but the material property in macroscopic conformation for such structures was not fully clarified, in spite of the high strength and high toughness of the spider silk. An orb-weaver spider has seven types of secretory glands. The seven different glands secrete seven different silks with different compositions and properties for different uses. The dragline is a lifeline for the spider and also used for the frame and radii of an orb web. The dragline and radial line form the structural skeleton, and the spiral lines are woven on the structural skeleton. Only the spiral lines can capture the prey because only the spiral line has a sticky ball coating.

\begin{figure*}[h]
\begin{minipage}[t]{17.8cm}
\resizebox{1\textwidth}{!}{%
  \includegraphics{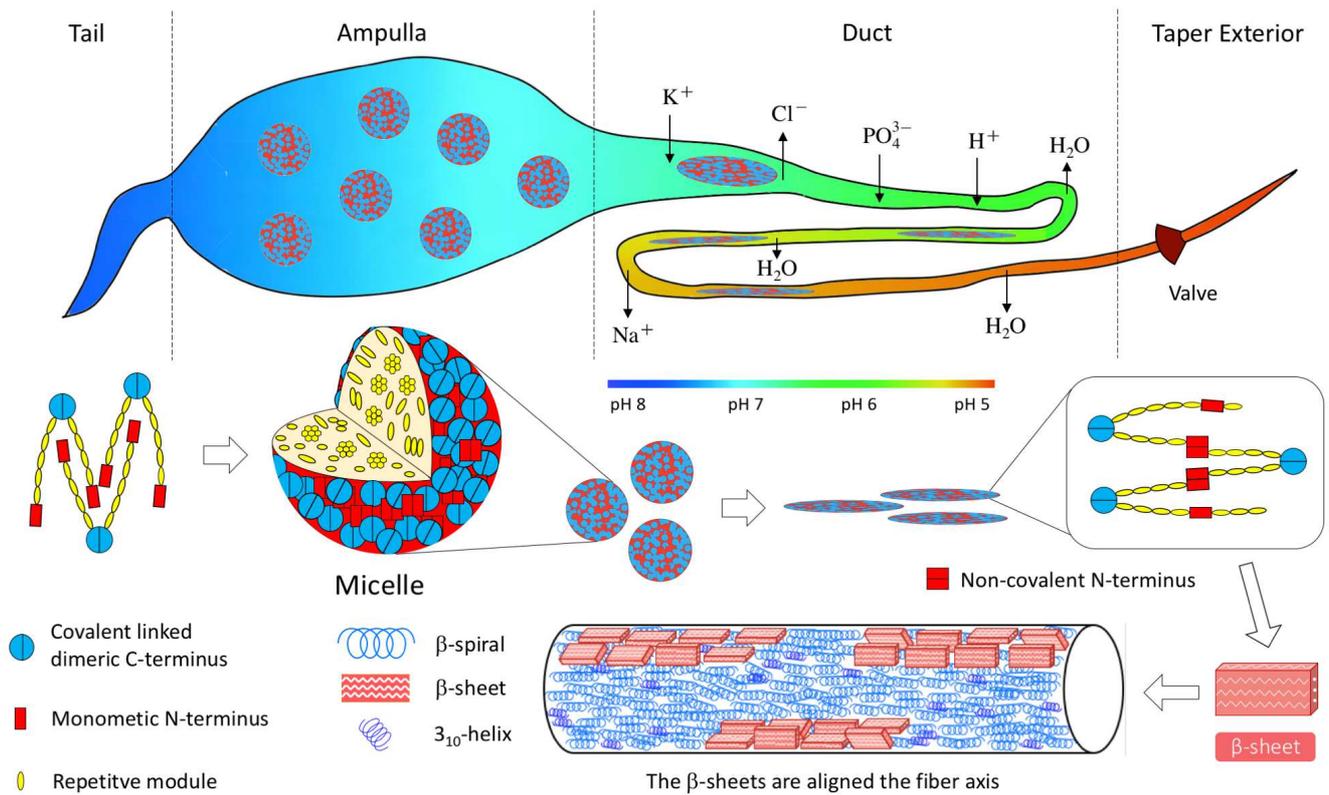}}
\end{minipage}%
\caption{Schematic of the spiders spinning apparatus and structural hierarchy in silk assembling related to assembly into fibers \cite{g38,g37,Zhao2017,g36,g30}.}
\label{fig1}      
\end{figure*}

Among them, the dragline has high strength and the radial line gives the strength of the skeleton constituting the orb web. The high strength characteristic of the dragline and the radial line is given by the \textbeta -sheet structure \cite{g8-39,g8-47} of fibroin molecules \cite{g3-30-13,g3-31,g3-20,g12,g3-30,g12-41,g12-39}.  The dragline and radial line are secreted by the ampullate gland inside the spider's body.  The ampullate gland is composed of four parts as shown in Fig. \ref{fig1}\cite{g38,g37,Zhao2017,g36,g30}.  In the process of dragline production, the primary structure protein is secreted first from secretory granules in the tail \cite{g8}.  In the ampullate (neutral environment, pH = 7), the proteins form a soft micelle of several tens of nanometers by self-organization because the hydrophilic terminals are excluded \cite{g3-29}.  In ampullate, the concentration of the protein is very high \cite{g8-47,g36-24}.  Then, the micelles are squeezed into the duct.  The long axis direction of the molecules is aligned parallel to the duct by a mechanical frictional force and partially oriented \cite{g8,g3-29,g3-16}.  The continuous lowering of pH from 7.5 to 8.0 in the tail to presumably close to 5.0 occurs at the end of the duct \cite{g38,g38-9,g38-13}.  Ion exchange, acidification, and water removal all happen in the duct \cite{g37}.  The shear and elongational forces lead to phase separation \cite{g37}.  In the acidic bath of the duct, the molecules attain a high concentration liquid crystal state \cite{g8-49}.  Finally, the silk is spun from the taper exterior. The molecules become more stable helixes and \textbeta -sheets from the liquid crystal.

The second-order nonlinear optical effect is sensitive to highly oriented structures \cite{BNO,12}.  In our previous report \cite{Zhao2017}, the dragline and the radial line have been found to show an optical second-harmonic generation (SHG) response.  However, no SHG signal was observed from the spiral line. Because the only difference between the constituents in the dragline, radial line, and the spiral line is the presence and absence of the \textbeta -sheets of the protein, and only the spiral line does not contain \textbeta -sheets, the origin of the SHG of the radial line and the dragline was inferred to be the oriented \textbeta -sheets.  The SHG intensity of the dragline was found to depend on the polarization of the incident light. An anisotropic structure is the origin of the SHG. As shown in the bottom of Fig. \ref{fig1}, a model \cite{Zhao2017} was proposed that the randomly rotating arrangement around the fiber axis of the \textbeta -sheets should give the anisotropy.

On the other hand, we hypothesize that the molecules of the dragline may be further aligned by an external force.  The SHG from the dragline should respond to the changes of the molecular alignment.  Hence, in this study, we measured the change of the SHG response by applying an external force to the dragline in the fiber axis direction.

\section{Method and Sample}\label{Method and Sample}

The collection of the sample and the observation method for observing SHG were the same as that described in our previous report \cite{Zhao2017}.  The excitation light pulse energy density was 16 \textmu J/mm$^2$.  We confirmed that the breaking of the draglines in Figs. \ref{fig2} and \ref{fig3} is not caused by the incident light.  Namely, no damage in draglines was observed after 1 hour irradiation with this energy density.  In the tensile testing as shown in Fig. \ref{fig2}(a), a dragline was pulled and stretched by 1 mm steps from the natural length until it finally breaks.   After the onset of the breaking, the dragline is separated into several thinner fibers and some of them broke first.  One measurement in Fig. \ref{fig2}(b) took less than 10 minutes, and the one in Fig. \ref{fig3} less than 30 minutes.

\section{Results}\label{Results}

\begin{figure}[h]
\centering
\begin{minipage}[t]{8.3cm}
\resizebox{1\textwidth}{!}{%
  \includegraphics{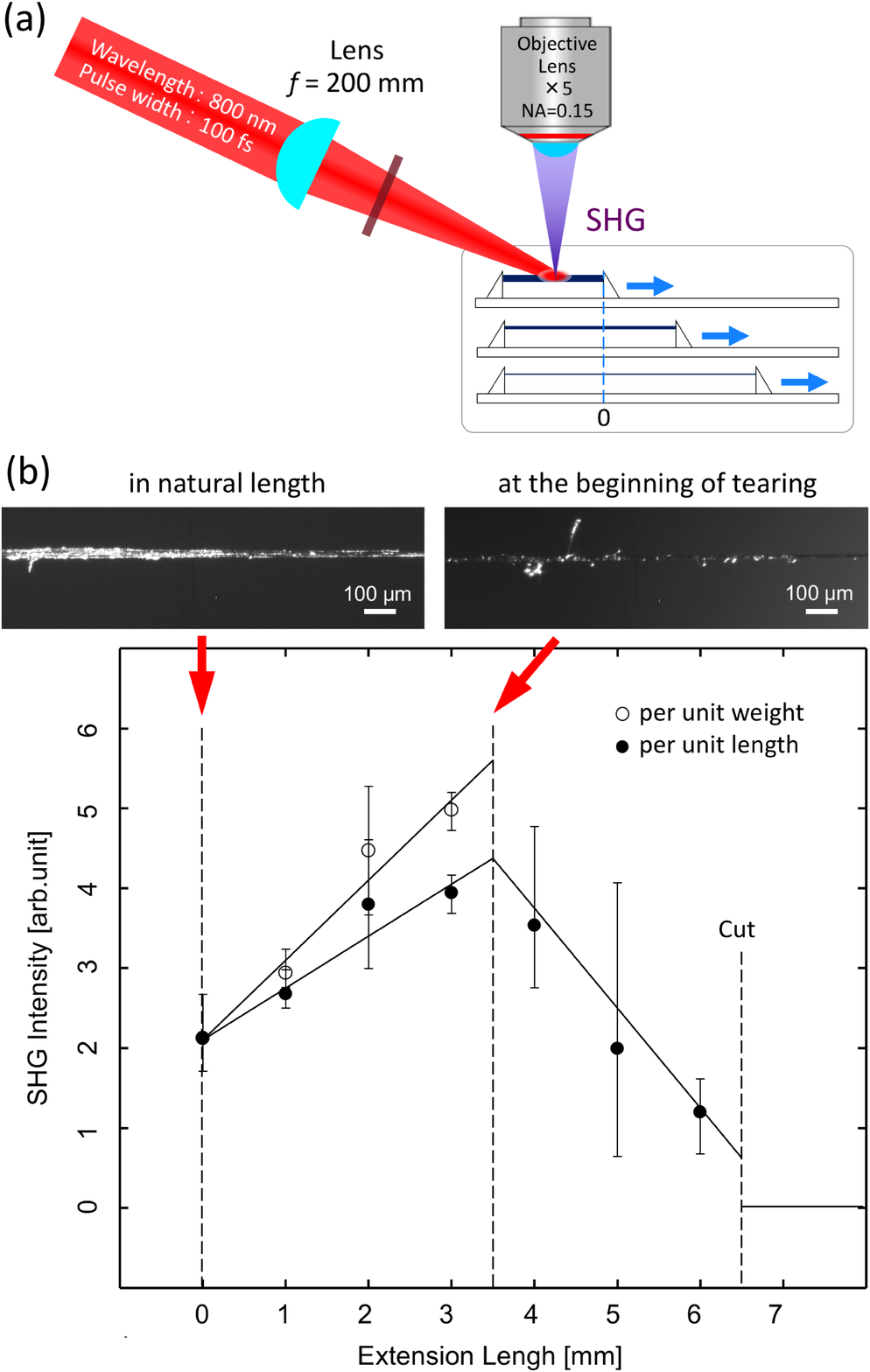}}
\end{minipage}%
\caption{(a) The second harmonic generation (SHG) observation method of a pulled dragline. A dragline was stretched by each 1 mm from the natural length (14.3 mm) until it broke down. The pulling speed was about 3$\times$10$^{-4}$ m/s.  The excitation light energy density of one pulse was 16 \textmu J/mm$^2$. (b) The relation of the SHG intensity and the length of the dragline. The top normal microscopic images are the dragline in the natural length and at the beginning of breaking stage of the dragline.}
\label{fig2}      
\end{figure}

As shown in Fig. \ref{fig2}(a), the dragline was pulled in its length direction from a natural state ({\it L}=14.3 mm).  The elongation length {\it \textDelta L} is defined as 0 in the natural state.  The length of the natural state is nearly equal to the one when the line is pulled by the force of the weight of the spider.  The dragline was pulled until it broke.  The incident light polarization was perpendicular to the fiber axis.  The polarization of the detected SHG was not chosen.  The SHG intensity was measured by microscope optics with an image intensified charge coupled device camera.  The intensity of the SHG became stronger with extension until a part of the dragline began to break at {\it \textDelta L}=3.4 mm.  The diameter of the dragline became thinner as it was stretched until it was broken apart.  The slope of SHG enhancement per unit mass as a function of {\it \textDelta L} is larger than that per unit length as shown in Fig. \ref{fig2}(b).  Breaking began when the dragline was stretched longer than {\it \textDelta L}=3.4 mm.  The microscopic images in Fig. \ref{fig2}(b) show a dragline in the natural length ({\it \textDelta L}=0) and at the beginning of breaking stage ({\it \textDelta L}=3.5 mm).  The dragline at the breaking position gradually became thinner and was cut.  The SHG weakened at the breaking part.  The total SHG intensity decreased while the dragline was broken continuously until it breaks apart.

\begin{figure}[h]
\centering
\begin{minipage}[t]{8.5cm}
\resizebox{1\textwidth}{!}{%
  \includegraphics{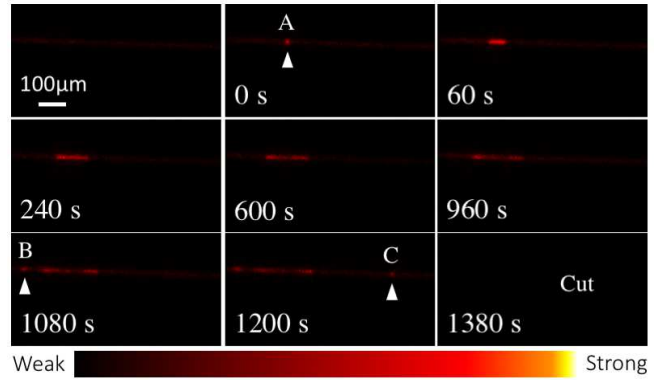}}
\end{minipage}%
\caption{Second harmonic generation microscopic images of a dragline while it is broken apart at a constant length. The time was defined as 0 second when a break point appeared.}
\label{fig3}      
\end{figure}

Next, we stretched the dragline until it began to break and stopped stretching at that length and observed the SHG image as a function of time. At a length near the breaking threshold, the dragline is slowly broken. A strong SHG spot A as shown in Fig. \ref{fig3} appeared at some point of time. The length of the strong SHG domain grew along the fiber axis from the initial location as the time elapsed. Then, a new strong SHG spot B appeared at time 1080 seconds. Another strong SHG spot C appeared at time 1200 seconds. Then the dragline broke at time 1380 seconds.

\section{Discussion}\label{Discussion}

\begin{figure}[h]
\centering
\begin{minipage}[t]{8.5cm}
\resizebox{1\textwidth}{!}{%
  \includegraphics{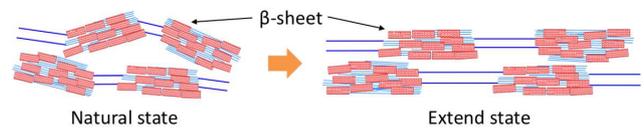}}
\end{minipage}%
\caption{Schematic representation of the effect of mechanical strain on the structure of a dragline. The natural state is a less ordered structure. The stretched state is a highly aligned structure.}
\label{fig4}      
\end{figure}

The \textbeta -sheets are aligned due to mechanical frictional force when passing through a thin duct in the spider's body and partial ordering occurs along the fiber axis \cite{g8,g3-29,g3-16,g36,g30}.  The origins of the elastic properties of dragline are similar to that of a synthetic elastomer \cite{g12-39}.  As shown in the model in Fig. \ref{fig4}, the flexible chain segments are stretched in the strained direction due to an external force and the hard crystal regions become more highly aligned. The orientation of the \textbeta -sheet becomes more unidirectional. The origin of the observed SHG enhancement effect is suggested to be the orientation of the \textbeta -sheet induced by stretching the dragline. From the linear relationship of the stress and strain of the dragline in previous reports \cite{g1,g41}, the result of Fig. \ref{fig2} is suggested to be a phenomenon seen in the stress increase stage before the breaking of the dragline.

The intensity change in Fig. \ref{fig2} for the extension length {\it \textDelta L} less than 3.5 mm is uniform and originates from an increase of the orientation of molecules in the whole dragline. On the other hand, Fig. \ref{fig3} is in the plastic deformation yield phase of the dragline slightly above the breaking threshold. Hence, we suggest that the SHG enhancements presented in Figs. \ref{fig2} and \ref{fig3} are from different origins. The phenomenon in Fig. \ref{fig3} can be explained by the following model. The alignment of the \textbeta -sheets in the stretched dragline is suggested to provide a stronger SHG intensity. The strong SHG spot A in Fig. \ref{fig3} was located at the mechanically weakest part in the dragline. The strong SHG intensity at spot A indicates well-aligned \textbeta -sheets.  There the \textbeta -sheets get oriented and the dragline becomes mechanically stronger and sustained the structure under the strain nearby the spot A. The domain of high orientation of the \textbeta -sheet extended as a function of time until 1080 seconds. From another report\cite{g40}, it is inferred that the helixes seen in the model figure at the bottom of Fig. \ref{fig1} broke first, while the \textbeta -sheet nanocrystals bore the strain until the dragline broke. When the part of the dragline near the spot A reaches the limit of flexibility, the \textbeta -sheets at other weak points B and C begin to be oriented one after another. The partial weakening of the SHG intensity in the A domain after 600 seconds in Fig. \ref{fig4} is due to partial breaking of the dragline. Eventually, the dragline broke near A because it could not withstand the stress. Namely, the highly oriented \textbeta -sheets had a function of prolonging the time until the dragline broke down by the pulling force.

\section{Summary\label{Summary}}

It has been discovered that the SHG response of the dragline of a spider is enhanced until it begins to break by a stretching force. The \textbeta -sheets in the dragline are suggested to be oriented more unidirectionally by the stretching force. After the onset of the breaking, the dragline is separated into several thinner fibers and some of them broke first. Here the SHG intensity integrated along the length direction decreased. We suggest that the highly unidirectional \textbeta -sheet orientation prolonged the time before breaking of the dragline. We also monitored the SHG image of the dragline as a function of time before it finally breaks down by the stretching force. After the onset of the breaking, the dragline is separated into several thinner fibers and some of them broke first by the stretching force. We found that the SHG intensity is enhanced at different positions at different times. In the dragline breaking, the \textbeta -sheets are suggested to begin to be highly oriented from the weakest location of the silk strength. This local orientation growth may be preventing the breaking of the dragline. The high strength of the dragline may be partly due to a physical structure of the \textbeta -sheets arranged in one direction.

\section*{Acknowledgements}

This research was financially supported by a JAIST Research Grant (Grant for Fundamental Research in 2017) from Japan Advanced Institute of Science and Technology Foundation (JAIST Foundation).

\bibliographystyle{0.zidingyi}
\bibliography{SIA}



\end{document}